\documentclass[twocolumn]{article}

\usepackage{authblk}
\usepackage{graphicx}

\usepackage{booktabs} 
\usepackage{framed} 
\usepackage{hyperref} 
\usepackage{multirow} 
\usepackage{soul} 
\usepackage{subcaption} 
\usepackage{tabularx} 
\usepackage{threeparttable} 
\usepackage{wasysym} 
\usepackage[table]{xcolor} 
\usepackage{xurl} 

\newenvironment{mybox}{\begin{shaded*}}{\end{shaded*}}

\definecolor{mybeige}{HTML}{DDC1B1}
\definecolor{myblue}{HTML}{5E62B8}
\definecolor{mypurple}{HTML}{C5C6E7}
\definecolor{mygray}{HTML}{878697}
\definecolor{mylightbeige}{HTML}{F5DCC9}
\definecolor{mylightpurple}{HTML}{E8E6FE}
\definecolor{shadecolor}{RGB}{247, 243, 239} 

\newcommand{\noo}{\Square}
\newcommand{\mbe}{\XBox}
\newcommand{\full}{\CIRCLE}
\newcommand{\half}{\LEFTcircle}
\newcommand{\emty}{\Circle}

\newcommand{\hlblue}[1]{{\sethlcolor{myblue!30}\hl{#1}}}
\newcommand{\hlpurple}[1]{{\sethlcolor{mypurple}\hl{#1}}}
\newcommand{\hlbeige}[1]{{\sethlcolor{mybeige!30}\hl{#1}}}
\newcommand{\hldarkbeige}[1]{\textcolor{white}{\sethlcolor{mybeige}\hl{#1}}}
\newcommand{\hldarkblue}[1]{\textcolor{white}{\sethlcolor{myblue}\hl{#1}}}
\newcommand{\hlgray}[1]{{\sethlcolor{mygray!30}\hl{#1}}}

\title{Making Cellular Networks Crisis-Proof: Towards Island-Ready, Resilient-By-Design 6G Communication Networks}
\author{Leon Janzen}
\author{Matthias Hollick}
\affil{Secure Mobile Networking Lab (SEEMOO), Technical University of Darmstadt, Germany}
\date{September 30, 2025}

\begin{document}

\maketitle




\begin{abstract} 
5G and 5G-Advanced cellular networks are vulnerable to regional outages resulting from disasters or targeted attacks.
This fragility stems from the reliance on the central core network involved for most 5G connectivity use cases.
Crisis-struck regions isolated from the cellular core network form \textit{islands}, where crisis response is hindered by the unavailability of recovery-relevant services, such as emergency calls, cell broadcasts, messengers, and news apps.
Our concept of island-ready, resilient-by-design 6G communication networks envisions local cellular connectivity allowing users to connect to regional application servers, which is currently impossible.
In our conceptualization, we follow an all-society approach, as realizing island connectivity requires the cooperation of multiple actors, including users, operators, developers, providers, and authorities.
We evaluate how island-ready 5G and 5G-Advanced systems are and outline the open challenges stakeholders must address for full island readiness, such as decentralizing the 6G core network and designing local-first application architectures.
\end{abstract}



\section{Introduction}
\label{sec: intro}

\begin{figure*}
\centering
\includegraphics[width=.75\linewidth]{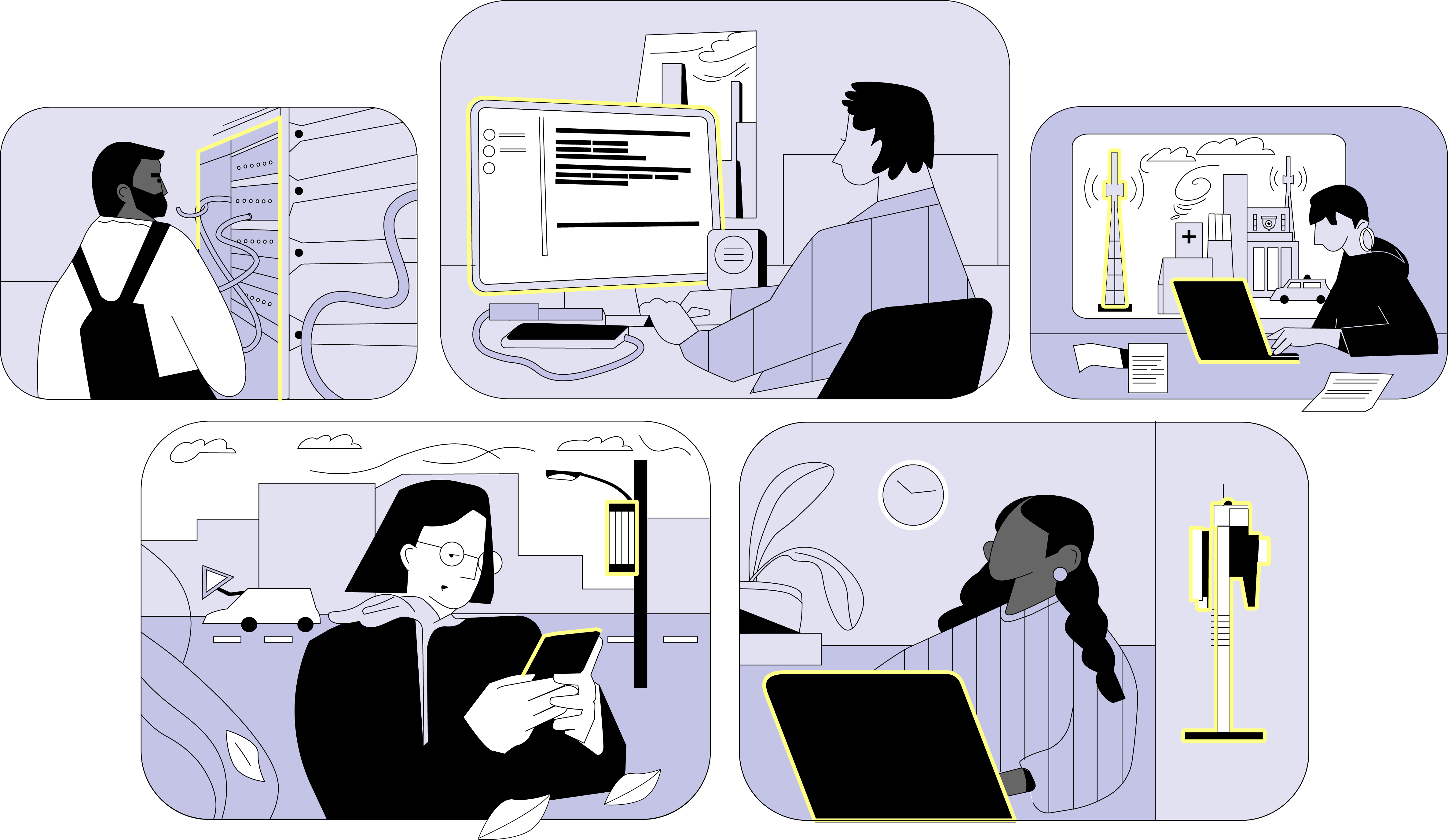}
\caption{%
    \textbf{The Vision of Island-Ready 6G Communication.}
}
\label{fig: vision}
\end{figure*}

Cellular networks are at the center of our modern digital world.
They connect mobile users around the globe, enabling them to communicate with each other and access websites and applications on the Internet.
Further, information and communication technologies are an integrated part of critical infrastructures, giving them a particular value during crisis situations.

5G systems are built around a core network, which is typically hosted in a central location with only a few gateways \cite{rula_2017_adopting}.
The circumstance that user traffic has to pass through the core network before going to the server, even if the server is hosted locally and the core network is hosted remotely, is called \textit{path inflation}~\cite{zarifis_2014_diagnosing} and has been mainly studied with a focus on performance~\cite{eismann_2016_collective,kiess_2014_centralized}.
However, the mechansims causing path inflation also severely impact the resilience of the communication system, as each link and system involved in connecting users to applications might fail~\cite{khaloopour_2024_resilience}.

Recent disasters and targeted attacks highlight the fragility of cellular networks.
For example, during the 2021 Western European flooding, water masses destroyed critical Internet infrastructure of the Ahr valley, leaving citizens without Internet access during crisis response~\cite{fekete_2021_here,mastd_2021_mobilfunk}.
Also, during the first days after the 2011 Great East Japan earthquake, cellular networks in the affected regions were overloaded and survivors reported poor connectivity~\cite{yamamura_2014_communication}.
The 2025 blackout on the Iberian Peninsula left 50 million people without electricity and led to the collapse of cellular networks, the regional Internet, and connected applications, such as electronic payment systems~\cite{gonzales_2025_social}.
Targeted attacks on transport network infrastructure, such as Internet exchange points~\cite{jain_2022_ukrainian,singla_2023_analysis} and undersea cables~\cite{guilfoyle_2022_final}, led to regional and nationwide outages in the past.

The loss of cellular connectivity drastically worsens crisis response in affected areas because many crisis-relevant smartphone services rely on it, for example, cell broadcasts, emergency calls, telephony, and text messaging~\cite{eismann_2016_collective,tan_2017_mobile}.
Additionally, without Internet connectivity, many crisis-relevant smartphone apps, such as messengers, news apps, crisis apps, and social media, are rendered useless~\cite{simko_2023_use}.
Traditional recovery mechanisms take time to set up, be it replacing broken links with directional antennas~\cite{suzuki_2013_directional}, deploying mobile base stations~\cite{trotta_2013_reestablishing}, crisis connectivity via satellites~\cite{volk_2021_emergency}, or drones~\cite{zhao_2019_uav}.
Yet, time is critical during the first hours of crisis response.
The recently introduced concept of \textit{island connectivity}~\cite{janzen_2025_user} proposes a cellular-native solution for communication recovery, outlining how users can access applications deployed at the network edge during crisis response, utilizing local cellular connectivity to connect smartphones and edge servers.

The central contribution of this article is the \textbf{analysis of island readiness} in the context of 6G communication networks. 
We introduce the corresponding definition in Section~\ref{sec: vision} based on resilience and resilience-by-design principles and extend the conceptual model of island connectivity by identifying stakeholders and their perspectives on island connectivity.
Figure~\ref{fig: vision} visualizes the collaboration of stakeholders in island-ready 6G communication networks.
We outline the main challenges stakeholders must solve to realize island-ready 6G communication.
To this end, we define and answer three research questions (RQ1-3) in this article: 

\begin{mybox}
    \textbf{RQ1:} Which stakeholders need to collaborate when realizing island-ready 6G communication?
\end{mybox}

Island connectivity involves various actors with different perspectives on the communication network.
Realizing the vision of island-ready 6G communication requires an all-society approach to encourage coordination of all actors \cite{clark_2025_strengthening}.
Section~\ref{sec: perspectives} identifies stakeholders involved in cellular communication networks and their perspectives.

\begin{mybox}
    \textbf{RQ2:} To which degree are 5G and 5G-Advanced cellular communication networks island-ready?
\end{mybox}

Island connectivity requires the deviation from traditional (mostly centralized) communication patterns because only systems hosted within the region are available during island scenarios.
To get an overview of the situation, Section~\ref{sec: architecture} considers today's communication network as a system of systems and evaluates the involved subsystems regarding their availability on islands.

\begin{mybox}
    \textbf{RQ3:} Which open challenges hinder actors in realizing/deploying island readiness?
\end{mybox}

The island readiness evaluation identifies a list of systems that are not yet island-ready.
We derive a roadmap with open challenges that 6G must solve to realize island connectivity.
Section~\ref{sec: roadmap} discusses what has to change to make these systems island-ready.

In summary, we contribute definitions surrounding island readiness, identify stakeholders and systems involved in communication networks, and evaluate the island readiness of 5G and 5G-Advanced systems.
We outline open challenges that involved actors have to solve, thus offering a research and development roadmap towards crisis-proof communication systems.

\section{Island-Ready Communication Networks}
\label{sec: vision}

This section evolves the conceptual model of island connectivity \cite{janzen_2025_user} to define island readiness as a multi-actor and system-of-systems~\cite{boardman_2006_system} problem.
We describe the vision of island-ready communication networks and introduce terminology to navigate the realm of island readiness.

\subsection{Communication Networks}

There are numerous definitions for \textit{communication networks}, e.g., for computer communication networks~\cite{frank_1971_computer}, with a mathematical interpretation~\cite{gomory_1964_synthesis} or an economic one~\cite{bolton_1994_firm}.
Our own socio-technological definition recognizes two dimensions of communication networks:

\begin{mybox}
A \textbf{communication network} is a multi-actor system of systems that enables connectivity between users acting as senders and receivers.
Communication involves numerous (sub-)systems under the control of multiple actors who have to cooperate to provide communication as an end-to-end service.
\end{mybox}

Many actors use the services provided by communication networks, e.g., users connecting to applications from their smartphones or developers making these applications accessible to mobile users.

\subsection{Island Connectivity}
Janzen et al.~\cite{janzen_2025_user} introduced the definitions for islands and island connectivity that we extend in this article:
\begin{mybox}
"An \textit{island} is a crisis-struck area isolated from the outside Internet, preventing citizens from using smartphone apps that require an Internet connection to remote servers"
and
"\textit{island connectivity} enables users on an island to use the cellular network within the island's bounds, facilitating the use of cellular services (e.g., emergency calls and cell broadcasts) within the island and access to servers on the island's network edge (e.g., multi-access edge computing)."~\cite{janzen_2025_user}
\end{mybox}

As applications potentially need to apply changes when operating in island connectivity compared to global connectivity, the terms \textit{island operation} (vs.~normal operation) and \textit{island mode} (vs.~normal mode) help to indicate an application's state around island connectivity.
These definitions take a user-centric perspective and focus on the service users receive from communication networks.
In the remainder of this section, we use island connectivity as the foundation for definitions of resilience and island readiness.

\subsection{Resilience}
We closely follow the resilience definition of Khaloopour et al.~\cite{khaloopour_2024_resilience} and adapt it for the use case of island connectivity:

\begin{mybox}
A \textbf{resilient} 6G communication network is prepared to face connection failures between regions, withstand them, and prevent most failures from causing performance degradation.
It can also absorb the isolation of regions into islands, ensuring island connectivity.
Moreover, resilient 6G communication networks can transition into and out of island operation, maintain state in island operation, and evolve based on the experiences learned during this process.
\end{mybox}

Hence, our resilience definition focuses on the service 6G provides to users, i.e., communication.
Because communication is a critical infrastructure in many countries~\cite{bbk_2024_informationstechnik,cisa_2024_communications}, making 6G resilient should be a top priority for all stakeholders.

\subsection{Island Readiness}

We use the above definitions of communication networks, islands, island connectivity, and resilience to define island readiness:

\begin{mybox}
  A 6G communication network is \textbf{island-ready} if resilience capabilities at different system levels and from different perspectives have been integrated into the system, enabling it to seamlessly provide island connectivity after the isolation of regions into islands.
\end{mybox}

Our definition of island readiness is close to the \textit{resilience-by-design} definition of Khaloopour et al.~\cite{khaloopour_2024_resilience}, adapted to the use case of island connectivity.
Arguably, island-ready 6G communication networks are, by design, prepared to cope with the isolation of regions.
However, it is possible to design resilient 6G communication networks that require small adjustments, such as the activation of standby replicas, for the (non-seamless) transition to island operation.
We consider such systems to be resilient but not island-ready.
Section~\ref{sec: architecture} uses the definitions of resilience and island readiness to evaluate the extent to which 5G-based communication networks can provide island connectivity.

\section{Stakeholders and Their Use Cases}
\label{sec: perspectives}

\begin{figure*}
\centering
\includegraphics[scale=0.4]{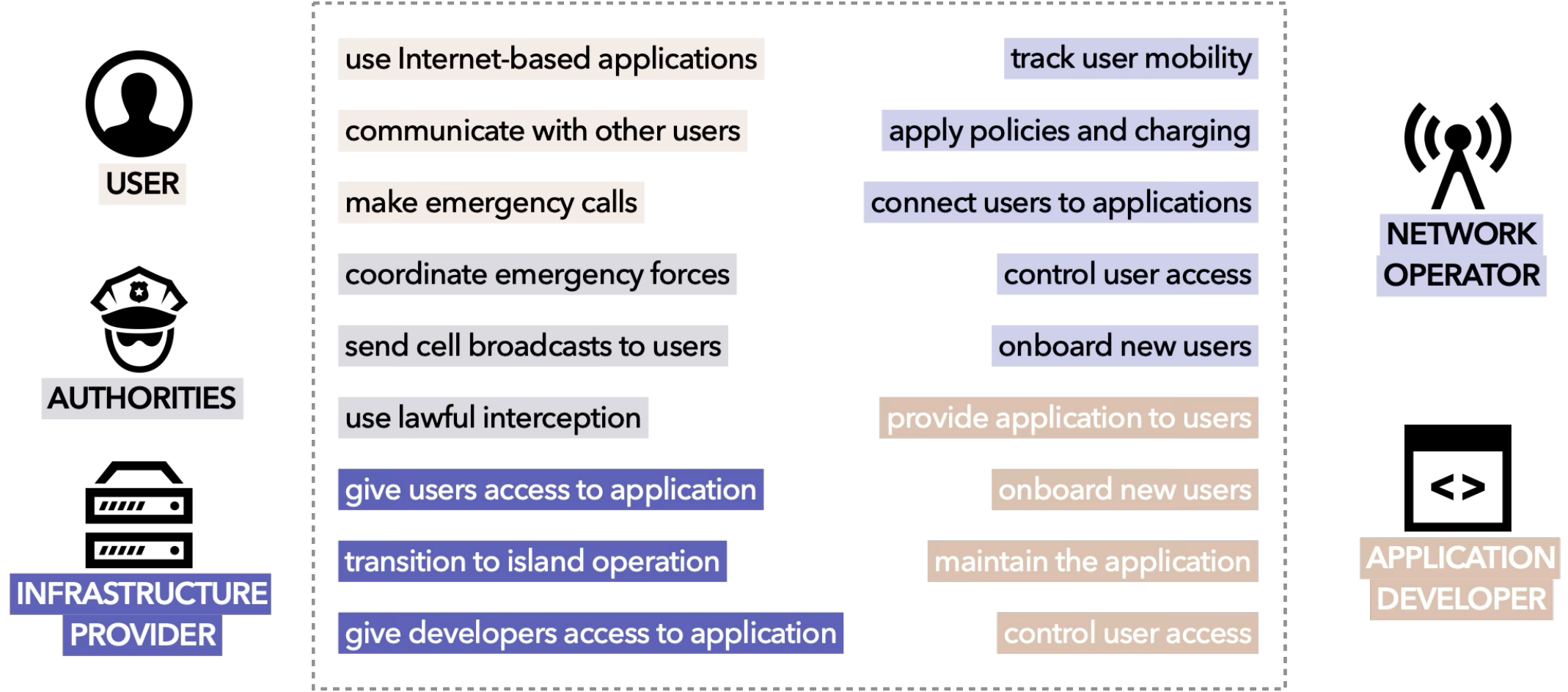}
\caption{%
    \textbf{Use Cases of Island-Ready 6G Communication.}
    Use cases are drawn in boxes, and the involved actors are represented as symbols around the frame.
}
\label{fig: usecases}
\end{figure*}

Billions of users worldwide take access to the Internet via cellular networks at all times for granted.
However, using 5G connectivity on a smartphone for apps that connect to the Internet is a complex process involving many systems.
Realizing island connectivity requires the operators of these systems to collaborate.
This section aims to identify stakeholders involved in today's communication networks and collect use cases from their perspectives (Figure~\ref{fig: usecases}).
Note that this collection is non-exhaustive and meant to give a first idea of which requirements island-ready communication has to support.
It requires in-depth expertise in the discussed perspectives to identify and refine additional use cases.

\subsection{The User Perspective}

\hlbeige{Users of communication networks} rely on everyday smartphone apps during crisis response and prefer using them in island scenarios over switching to dedicated crisis apps~\cite{janzen_2025_user}.
The primary needs of users in island scenarios are to make emergency calls and communicate with other users by telephone or text messaging.
Besides, it would be beneficial if users could use smartphone applications during crisis response, e.g., messengers, social media, and news or crisis apps~\cite{lukau_2024_public}.
We use the term \textit{user} to refer to humans affected by the crisis and using the services of 6G communication networks.

\subsection{The Operator Perspective}
The \hlpurple{mobile network operator} is operating the cellular core network, an access network, and, including telephony and emergency calls in the picture, also an IP multimedia subsystem (IMS).
They hand out subscriber identity modules (SIMs) to onboard new users and store the corresponding identifier and cryptographic keys to control user access.
Operators connect to user equipment through an access network to track user mobility and forward calls and data to them.
They connect users to applications on the Internet or other connected data networks, such as the IMS for telephony, text messages, and emergency calls.
Finally, they apply policies and charging corresponding to the user's subscription.

\subsection{The Developer Perspective}
Service providers and \hldarkbeige{application developers} design, deploy, and operate applications on the Internet to be accessible to users.
The way apps and websites are accessed is similar, and both provide a service to users, so we generalize apps and websites under the term \textit{application}.
Developers onboard new users and maintain customer data to control user access to the application.
One decision developers have to make is where to deploy their application server, e.g., self-managed on-premise, web hosting, using the services of a cloud provider, or using the services of a cloud and edge provider.
Moving applications closer to the user has gained popularity over the last decade~\cite{oyeniran_2024_comprehensive}, introducing cloud and edge providers as stakeholders whose perspective we discuss next.
We use the term application developer or simply \textit{developer} to refer to both service providers and application developers and distinguish them from cloud and edge providers more easily while maintaining readability.

\subsection{The Provider Perspective}
We refer to \hldarkblue{cloud and edge providers} who offer web hosting, content delivery networks, or multi-access edge computing on cloud and edge infrastructure as \textit{providers}.
Developers use a provider's service to host their application and make it accessible to users.
As such, the prominent use case of providers is to make an application available to users.
Besides, providers maintain an interface to enable developers to access their application, e.g., for fixes, updates, and retrieving data.
Providers must be ready to transition into island operation, so users and developers can access the hosted applications on islands.

\subsection{The Authorities' Perspective}
For our crisis definition, island scenarios involve one or multiple agencies.
\hlgray{Authorities} for civil protection warn users via cell broadcasts during natural disasters~\cite{aloudat_2011_application}.
Law enforcement agencies can wiretap users via lawful interception, which might be relevant in island scenarios caused by dedicated cyber attacks or armed conflicts~\cite{3gpp_33_126}.
In some countries, emergency services such as police forces, the fire department, and ambulances use public safety networks based on the cellular network~\cite{savunen_2023_role}.

RQ1 seeks actors whose perspective needs consideration when realizing island-ready 6G communication, which we answer as follows:

\begin{mybox}
\subsection*{Stakeholders of Cellular Communication Networks}
This section started from the users and iteratively discovered relevant stakeholders.
\textbf{Users} rely on cellular communication networks, especially in crisis scenarios.
\textbf{Operators} offer connectivity as a service to users and need to adhere to the authorities' regulations.
\textbf{Developers} deploy applications on the Internet for users to access.
\textbf{Providers} maintain cloud and edge capabilities to host applications.
\textbf{Authorities} retain communication means for crisis communications.
\end{mybox}

Our interpretation of communication networks recognizes the multiple stakeholders involved in providing cellular connectivity.
The actors in this section maintain the systems discussed in Sections~\ref{sec: architecture}~and~\ref{sec: roadmap}.

\section{Island Readiness of 5G Communication Networks}
\label{sec: architecture}

\begin{figure*}
	\centering
	\includegraphics[width=.75\linewidth]{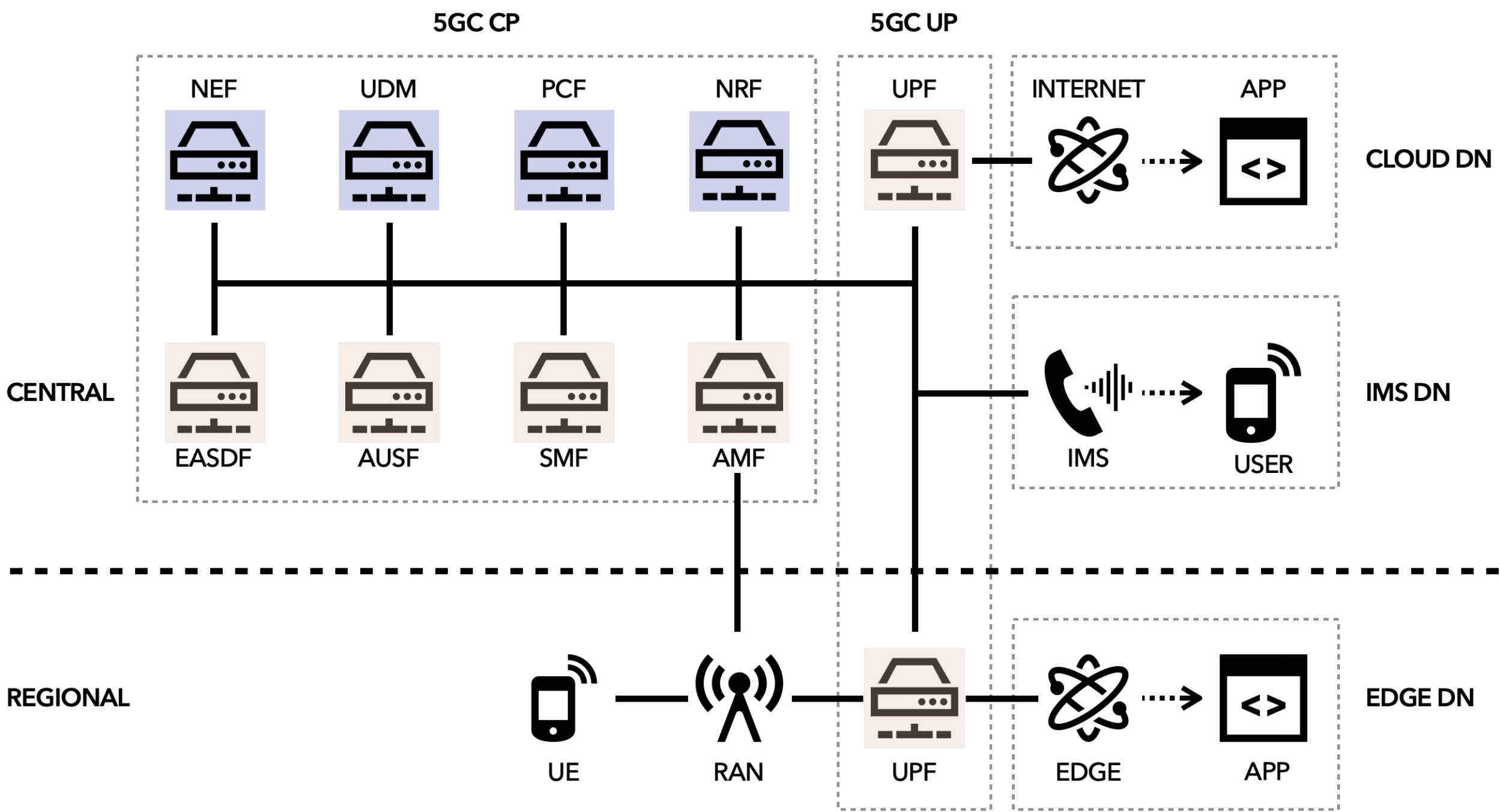}
	\caption{%
        \textbf{System-of-Systems View of Cellular Communication Networks.}
        This figure illustrates the subsystems of a 5G-Advanced communication network, providing a closer look at the control plane (CP) and user plane (UP) of the 5G core network (5GC).
        \hlbeige{Stateless core functions} and \hlblue{stateful core functions} are highlighted with background color.
        The horizontal dotted line separates systems deployed regionally from those deployed remotely.
        In an island scenario, only regional systems would be available, which is why 5G-Advanced is not island-ready for most use cases.
    }
	\label{fig: system-of-systems}
\end{figure*}

5G-based communication networks comprise the 5G system and additional systems, e.g., the IMS, transport networks, cloud, and edge servers.
This section describes these systems and evaluates their island readiness.

\subsection{The Internet and Its Clouds and Edges}
The Internet is a global network of networks with systems communicating via the Internet Protocol (IP) suite~\cite{kenett_2015_networks}.
With hosts, servers, and routers all over the globe, the Internet is decentralized by design~\cite{moura_2020_clouding}.
Cloud and edge computing paradigms reflect server location categories with virtualization technologies, allowing for the flexible movement of services to the most suitable place.
However, ultimately, servers are bound to one or multiple physical locations and made available to users through IP traffic routed through physical transport networks.
Island connectivity violates the core assumption of global routability~\cite{gao_2001_stable}, on which much of the current infrastructure is built.
In an island scenario, only a fraction of the Internet's resources are available regionally, as some parts of the Internet are hosted locally within the region, while others are hosted remotely in other (disconnected) regions.
Island readiness requires fundamental rethinking of availability and routability on the Internet.

Edge applications typically capitalize on the proximity to the user for low-latency use cases or to offload compute-heavy tasks from mobile devices~\cite{antevski_2020_integration}.
However, edge applications can also serve other purposes, e.g., providing crisis-relevant smartphone apps like messengers, crisis or news apps~\cite{lukau_2024_public}.
Edge servers are already available in many regions with hyperscalers extending their infrastructure to be closer to the user~\cite{zolfaghari_2020_content}.

\subsection{Applications}
Developers deploy apps, websites, and services on servers on the Internet to be accessible to users.
Providers offer different deployment options for such servers with different levels of performance, scalability, and availability~\cite{oyeniran_2024_comprehensive}.
Deploying small applications in a single-server setup is often sufficient, so smaller applications are typically centralized.
Popular applications, like social media or messengers, have to scale to billions of users and run in massive data centers across multiple geographic regions~\cite{google_2025_cloud}.
While high-availability cloud deployments guarantee wide-area geo-redundancy, much more fine-grained distribution levels are necessary to make applications island-ready~\cite{jammal_2015_high}.
Applications are transformable to island-ready applications by adopting decentralized or federated architectures, but these changes might increase the complexity of application backends~\cite{oyeniran_2024_comprehensive}, thus impacting scalability.
Currently, only a few applications use the potential of edge deployments~\cite{antevski_2020_integration}.
Anticipating a trend of local-first applications, providers should extend their edge infrastructure to increase capacity to deploy applications on the edge, and developers should migrate to island-ready architectures.

\subsection{5G Core Network}
We base the evaluation on specifications of the 3rd Generation Partnership Project (3GPP) and differentiate 5G and 5G-Advanced, attributing Releases 15~-~17 to 5G and Releases 18~and~19 to 5G-Advanced.
Figure~\ref{fig: system-of-systems} depicts a 5G-Advanced communication network as a system of systems with an exploded view of the 5G core network.
The 3GPP describes the cellular system architecture in TS~23.501~\cite{3gpp_23_501} with the 5G core network at its center.
The core was redesigned from 4G to 5G, introducing several new design choices~\cite{3gpp_29_244}.
The 5G core natively supports multiple data networks (DNs), e.g., forwarding user traffic to the Internet, the IMS, or enterprise networks.
Also, it natively separates control plane and user plane functions:

\paragraph{User Plane}
5G and 5G-A support regional deployment of the \hlbeige{user plane function (UPF)}, enabling users a shortcut to the regional Internet.
However, user plane connections require an initial connection to the control plane~\cite{3gpp_23_502}, e.g., to the \hlbeige{access and mobility management function (AMF)}, rendering the regional user plane inoperable without the control plane.

\paragraph{Control Plane}
The 5G control plane features a service-based architecture, i.e., each core function offers services via APIs~\cite{3gpp_23_502}, which enables scaling and cloud-native deployments of the 5G core.
It spans a variety of dedicated core network functions that were designed to be cloud-native and can be deployed virtually anywhere.
Most functions are stateless with only a few managing state~\cite{3gpp_23_501}, which are highlighted in a blue background color in Figure~\ref{fig: system-of-systems}.
These are the \hlblue{unified data management (UDM)} for subscriber data management, the \hlblue{policy control function (PCF)} for policy rules, the \hlblue{network exposure function (NEF)} managing secure access of trusted applications to core network services, and the \hlblue{network repository function (NRF)} enabling core functions to locate other core functions.
The \hlbeige{edge application server discovery function (EASDF)} is a 5G core function that enables the discovery of applications hosted on the local network edge~\cite{3gpp_23_548,3gpp_23_558}.
The \hlbeige{session management function (SMF)} and a domain name system (DNS) server can interact to point users to the closest edge application server~\cite{3gpp_23_548}, preparing domain name resolution for island connectivity.

\subsection{IMS and Interconnect}
The IMS is specified in TS~23.228~\cite{3gpp_23_228}.
It provides telephony and multimedia services, text messaging, and emergency services.
In 3GPP terminology, the IMS lives in an external data network with users connecting to it with IP-based 5G user plane traffic via the user plane.
On the control plane, the IMS talks to the 5G core's policy control function for quality of service and to the unified data management and \hlbeige{authentication server function (AUSF)} for accessing subscriber data and authentication.
The IMS enables global interconnections to users in other networks, e.g., users of other operators or fixed broadband users.
These interconnections are complex and security-critical, requiring coordination with other operators.
With 5G, the IMS has been virtualized and supports cloud-native deployment~\cite{de_2023_multi}.
Like the control plane, the IMS consists of stateless and stateful parts.
While distributed deployment of the IMS is possible, most operators run their IMS centrally or slightly distributed with low geo-redundancy~\cite{de_2023_multi}.

\subsection{Access and Transport Networks}
The radio access network (RAN), specified in TS 38.401~\cite{3gpp_38_401}, transports IP and management messages between the user equipment (UE) and the 5G core.
The 5G core is access agnostic, i.e., it supports access networks utilizing different RATs such as 5G New Radio, LTE, or Wi-Fi.
Access networks are distributed by design, bridging the distance between users and the 5G core.
There are other types of access networks, and some might introduce additional systems, such as the central units and distributed units of the Open Radio Access Network (Open RAN)~\cite{polese_2023_understanding}.

The transport network consists of fiber cables and internet exchange points connecting subsystems and networks across regions ~\cite{xu_2004_properties}.
Mobile network operators and cloud providers often maintain their own transport networks, often with peering agreements to connect multiple transport networks and forward each other's IP traffic~\cite{gill_2008_flattening}.
Like the original Internet, transport networks are decentralized by design.

For RQ2, we evaluated the systems of 5G-based communication networks to see which systems are island-ready.

\begin{mybox}
\subsection*{Island Readiness of 5G Communication Networks}

\textbf{The Internet is ready for island connectivity.}
The decentralized design of the Internet and the IP suite can seamlessly transition to island connectivity.
However, today's Internet relies on centralized and hierarchical services, such as DNS, which might need adaptation or reconfiguration for island readiness.

\textbf{Today's applications are not ready for island connectivity.}
The primary use cases of edge computing are related to low-latency communication or computation offloading.
While providers offer high-availability deployments, these often only include hosting the application on a few geo-redundant cloud servers, which is a step in the right direction but not enough to make crucial applications island-ready.

\textbf{Access networks and transport networks are ready for island connectivity.}
Access and transport networks are distributed by design, with physical networks spanning the globe.

\textbf{5G and 5G-Advanced systems are not ready for island connectivity.}
While the 5G user plane and many functions of the control plane are easily distributable, the few stateful core functions (UDM, PCF, NEF, and NRF) make it challenging to realize regional availability of the 5G core.

\textbf{The IMS is not ready for island connectivity.}
The IMS is cloud-native but features a few functions with stateful connections to 5G core functions or other IMSes of other operators.
\end{mybox}

Nevertheless, the central deployment of the 5G core and IMS is less a result of the impossibility of distributing stateful functions but more a result of the task's complexity and the lack of reward for operators to implement it.

\section{Open Challenges Towards Island Readiness}
\label{sec: roadmap}

\begin{table*}
\caption{%
    \textbf{Actors, Use Cases, and Island Readiness of 5G and 5G-A Communication Networks}
    From left to right, the columns illustrate the stakeholders of island-ready communication, their respective use cases, the involvement of the 5G core network (5GC) user plane (UP) and control plane (CP), the IP multimedia subsystem (IMS), and edge availability to support these use cases.
    On the right, it shows the island \hlbeige{readiness of 5G and 5G-Advanced standards} and what \hlblue{selected open challenges} are left for 6G to solve.
    }
\label{tab: use cases}
\centering
\scriptsize
\begin{threeparttable}
\begin{tabularx}{\linewidth}{llcccc>{\columncolor{shadecolor}}c>{\columncolor{shadecolor}}cX}
\toprule
\textbf{Actor}
& \textbf{The actor wants to \dots}
& \rotatebox{90}{\textbf{Core UP}}
& \rotatebox{90}{\textbf{Core CP}\tnote{a}}
& \rotatebox{90}{\textbf{IMS}}
& \rotatebox{90}{\textbf{Edge}}
& \rotatebox{90}{\textbf{5G}}
& \rotatebox{90}{\textbf{5G-A}}
& \textbf{Enabling this use case requires \dots}
\\
\midrule 
\multirow{3}*{User} 
  & communicate with other users.\tnote{b}        & \full & \half & \full & \emty & \noo & \noo & \hlblue{decentralized 6G-CP} and \hlblue{decentralized IMS} \\
  & make emergency calls.                       & \full & \half & \full & \emty & \noo & \mbe & \hlblue{decentralized IMS} \\
  & use Internet-based applications.            & \full & \full & \emty & \full & \noo & \noo & \hlblue{decentralized 6G-CP} and \hlblue{local-first apps} and \hlblue{edge capacity} \\
\midrule
\multirow{5}*{Operator} 
  & onboard new users.                          & \emty & \full & \emty & \emty & \noo & \noo & \hlblue{decentralized 6G-CP} \\
  & track user mobility.                        & \emty & \full & \emty & \emty & \noo & \noo & \hlblue{decentralized 6G-CP} \\
  & control user access.                        & \emty & \full & \emty & \emty & \noo & \noo & \hlblue{decentralized 6G-CP} \\
  & connect users to applications.              & \full & \full & \emty & \emty & \noo & \noo & \hlblue{decentralized 6G-CP} and \hlblue{local-first apps} and \hlblue{edge capacity} \\
  & apply policies and charging.                & \emty & \full & \emty & \emty & \noo & \noo & \hlblue{decentralized 6G-CP} \\
\midrule
\multirow{3}*{Developer} 
  & onboard new users.                          & \full & \half & \emty & \full & \noo & \noo & \hlblue{decentralized 6G-CP} and \hlblue{local-first apps} and \hlblue{edge capacity} \\
  & provide application to users.               & \full & \half & \emty & \full & \noo & \noo & \hlblue{decentralized 6G-CP} and \hlblue{local-first apps} and \hlblue{edge capacity} \\
  & control user access.                        & \full & \half & \emty & \full & \noo & \noo & \hlblue{decentralized 6G-CP} and \hlblue{local-first apps} and \hlblue{edge capacity} \\
  & maintain the application.                   & \full & \half & \emty & \full & \noo & \noo & \hlblue{decentralized 6G-CP} and \hlblue{local-first apps} and \hlblue{edge capacity} \\
\midrule
\multirow{2}*{Provider} 
  & give users access to application.           & \full & \half & \emty & \full & \noo & \noo & \hlblue{decentralized 6G-CP} and \hlblue{edge capacity} \\
  & give developers access to application.      & \full & \half & \emty & \full & \noo & \noo & \hlblue{decentralized 6G-CP} and \hlblue{edge capacity} \\
  & transition to island operation.             & \full & \half & \emty & \full & \noo & \mbe & \hlblue{edge capacity} and \hlblue{available staff} \\
\midrule
\multirow{3}*{Authorities}  
  & send cell broadcasts to users.              & \full & \full & \emty & \emty & \noo & \noo & \hlblue{decentralized 6G-CP} \\
  & coordinate emergency forces.                & \full & \half & \emty & \half & \noo & \mbe & \hlblue{decentralized 6G-CP} and \hlblue{island-ready regulations} \\
  & use lawful interception.\tnote{c}           & \full & \full & \half & \half & \noo & \noo & \hlblue{decentralized 6G-CP} and \hlblue{island-ready regulations} \\
\bottomrule
\end{tabularx}
\begin{tablenotes}
  \item A system is fully (\full), partly (\half), or not (\emty) involved in a use case.
        5G and 5G-A are partly (\mbe) or not (\noo) island-ready for a use case.
  \item[a] In this column, \half~indicates that the control plane is only required to initiate communications but not for subsequent connections.
  \item[b] Communication refers to telephony and text messaging.
  \item[c] In this row, the IMS and Edge are marked as \half~because without them, there is little to intercept.
\end{tablenotes}
\end{threeparttable}
\end{table*}

Table~\ref{tab: use cases} revisits the use cases identified in Section~\ref{sec: perspectives} and shows which systems of the cellular communication network are required for the use cases.
5G and 5G-Advanced lack support for most use cases without modifications.
We outline the main challenges towards island readiness.

\paragraph{Decentralization of Stateful Systems}
Most use cases require regional control plane availability, which 5G-Advanced does not support.
Operators need to find ways to distribute and \hlblue{decentralize the 6G control plane and the IMS} without compromising security and privacy.
While distributing stateless functions is simple, as they are cloud-native and can be deployed anywhere, distributing the stateful functions is hard because it requires synchronization of critical data~\cite{de_2021_distributed}.

\paragraph{Local-First Applications}
Many applications are hosted in a single or a few data centers, with scalability being the primary driver to deploy multiple replicas of the same application~\cite{oyeniran_2024_comprehensive}.
While providers offer high-availability deployments, they are insufficient to achieve the widespread and regional geo-redundancy required for island readiness~\cite{moreno_2018_orchestrating}.
Developers should find ways to \hlblue{design local-first application designs}, which will likely be infeasible for some applications or come at the expense of reduced application functionality.
However, island-relevant applications should be a priority~\cite{janzen_2025_user}.

\paragraph{Edge Capacity}
Realizing island connectivity shifts part of the storage and compute hotspots to the edge of the Internet.
Demand for edge deployments is growing, but still low compared to the envisioned edge demand of island-ready applications~\cite{haibeh_2022_survey}.
Providers should build \hlblue{regional edges with sufficient capacity} for crisis-relevant applications to prepare for island connectivity.

\paragraph{Manual Actions to Enable Island Operation}
Once a region becomes an island, communication into and out of the island is difficult.
If developers want to make their application island-ready, it should be able to seamlessly transition into island operation.
Otherwise, the application would experience downtime until necessary manual actions are taken to power the application in island operation.
In some cases, \hlblue{qualified staff needs to be in the region}, which might be hard, if not infeasible.

\paragraph{Island-Ready Regulations and Agreements}
Regulations serve as a common ground for all actors.
In crisis situations, different agreements can apply, e.g., operators are required to accept emergency calls from all users, including users from other operators or without a valid SIM card~\cite{3gpp_23_167}.
Scaling up similar infrastructure sharing for general crisis operation would be highly desirable.
Island readiness aims for the 6G communication network to be prepared for island scenarios by design, which requires fundamental changes to the cellular network and applications.
Making these systems island-ready requires the involved actors to \hlblue{issue regulations and agreements compatible with island readiness} before crises.

RQ3 seeks open challenges to be solved to realize 6G island readiness, and we selected and outlined a few of them.

\begin{mybox}
\subsection*{Roadmap Towards Island-Ready 6G}
\textbf{Operators} need to find ways to distribute stateful systems, such as the 6G core network and the IMS.
\textbf{Developers} need to make their applications available on a regional level.
\textbf{Providers} need to extend their edge capacities to prepare for the run on regional deployments.
\textbf{Authorities} need to support endeavors towards island-readiness with island-compatible regulations.
\end{mybox}

\section{Discussion}
\label{sec: discussion}

\subsection{The User in the Loop}
The user perspective on island connectivity was studied among adult smartphone users from major German cities in ~\cite{janzen_2025_user}.
The results indicate strong support for island readiness, but some participants were confused about the scope and limits of island connectivity.
A few participants feared that island-ready communication networks could facilitate forced isolation of regions, e.g., by the government.
These concerns should be addressed to facilitate widespread acceptance of island connectivity.
Contributions from the area of crisis communication can help, such as usability studies of device-to-device messengers~\cite{haesler_2021_connected,khaled_2019_case} that also had to deal with new connectivity models.

\subsection{Machine-to-Machine Communication}
This work is motivated from a user-centric perspective because island connectivity can save lives during crisis response.
However, there are also crisis-relevant use cases that follow a machine-to-machine communication pattern, e.g., automatically reading sensor data to get an overview of the situation~\cite{di_2024_emerging} or controlling cyber-physical infrastructure systems in reaction to sensed events.
Island connectivity can facilitate the continuous access of 5G/6G-enabled Internet of Things (IoT) deployments that would otherwise not be possible.
Again, this requires the developer of the corresponding applications to design them in an island-ready fashion.

\subsection{Implications on Security and Privacy}
Considering security aspects in island connectivity might introduce additional challenges, e.g., the applicability of end-to-end encryption in messengers~\cite{alvarez_2016_maintaining} or the privacy implications of knowing which application server is closest to a user~\cite{redact_2025_high}.
Additionally, providing secure decentralized operation poses a number of challenges over today's security architectures~\cite{kim_2010_secure}.
Researchers, operators, and authorities need to be able to quantify the resilience of both island-ready as well as non-island-ready communication networks.
This facilitates systematic evaluation of crisis operation capabilities, e.g., using stress tests or real-world exercises.

\subsection{Application Deployment Strategies}

\begin{figure*}
    \centering
    \begin{subfigure}[t]{.48\linewidth}
         \centering
         \includegraphics[width=\linewidth]{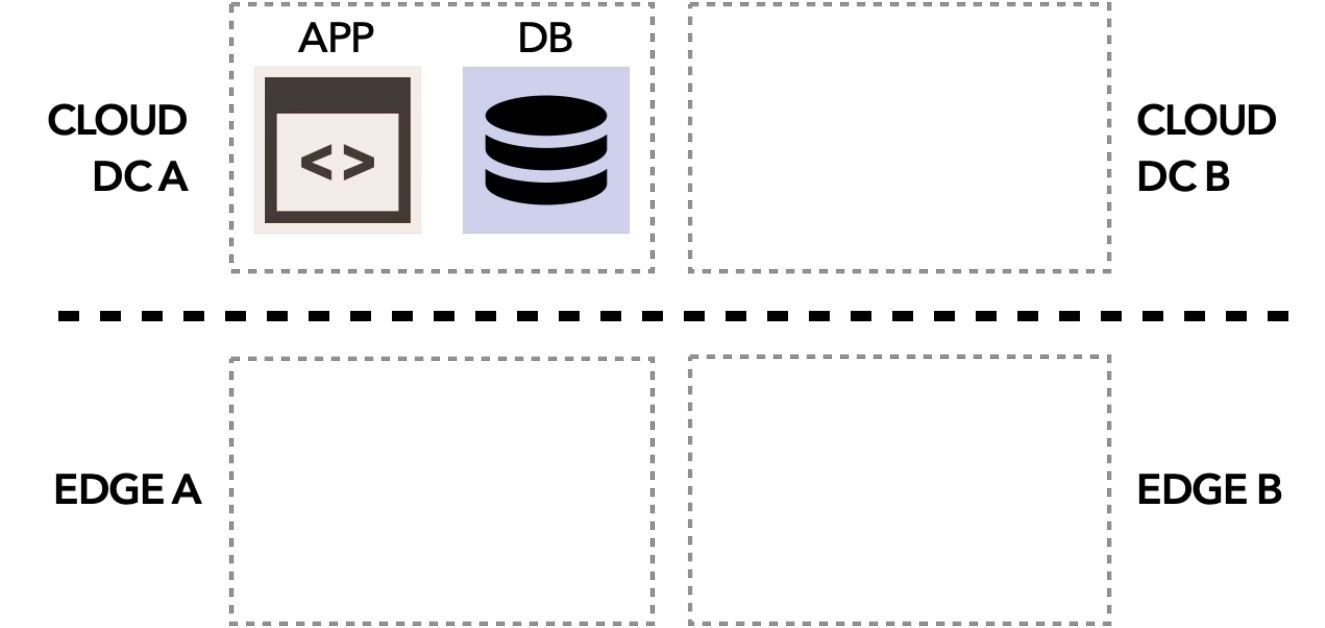}
         \subcaption{%
            Centralized Deployment
         }
         \label{fig: deployment centralized}
    \end{subfigure}
    \hfill
    \begin{subfigure}[t]{.48\linewidth}
         \centering
         \includegraphics[width=\linewidth]{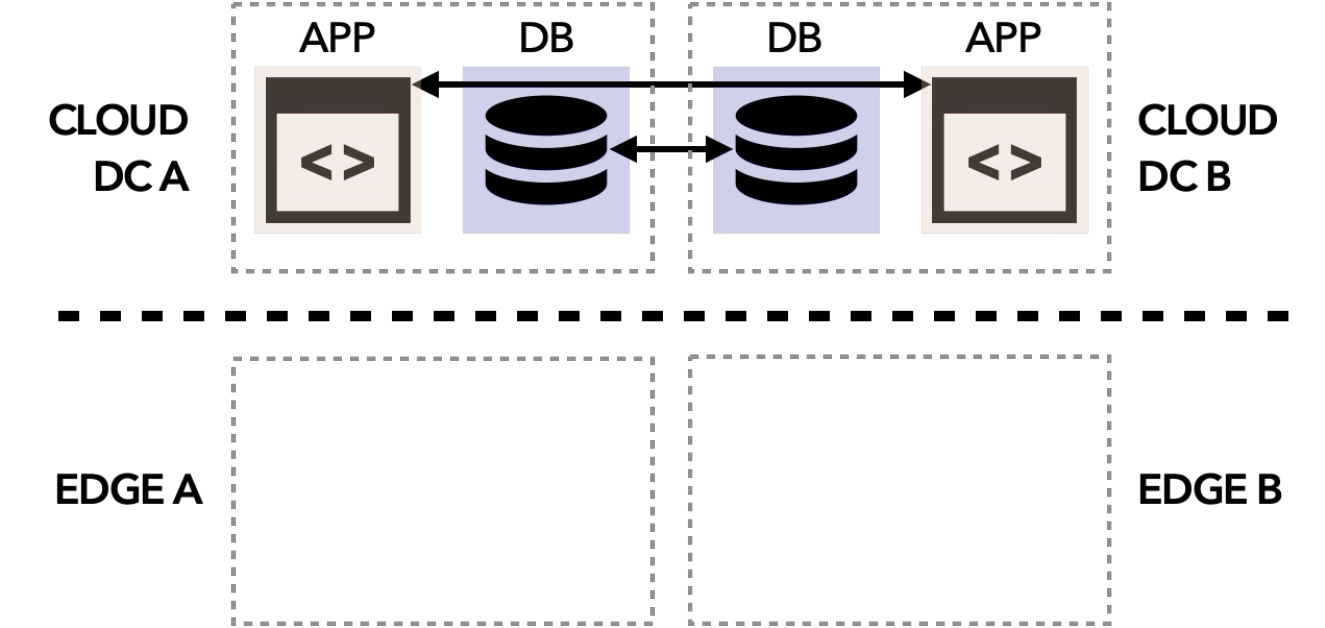}
         \subcaption{%
            High-Availability Deployment
         }
         \label{fig: deployment high-available}
    \end{subfigure}
    \begin{subfigure}[t]{.48\linewidth}
         \centering
         \includegraphics[width=\linewidth]{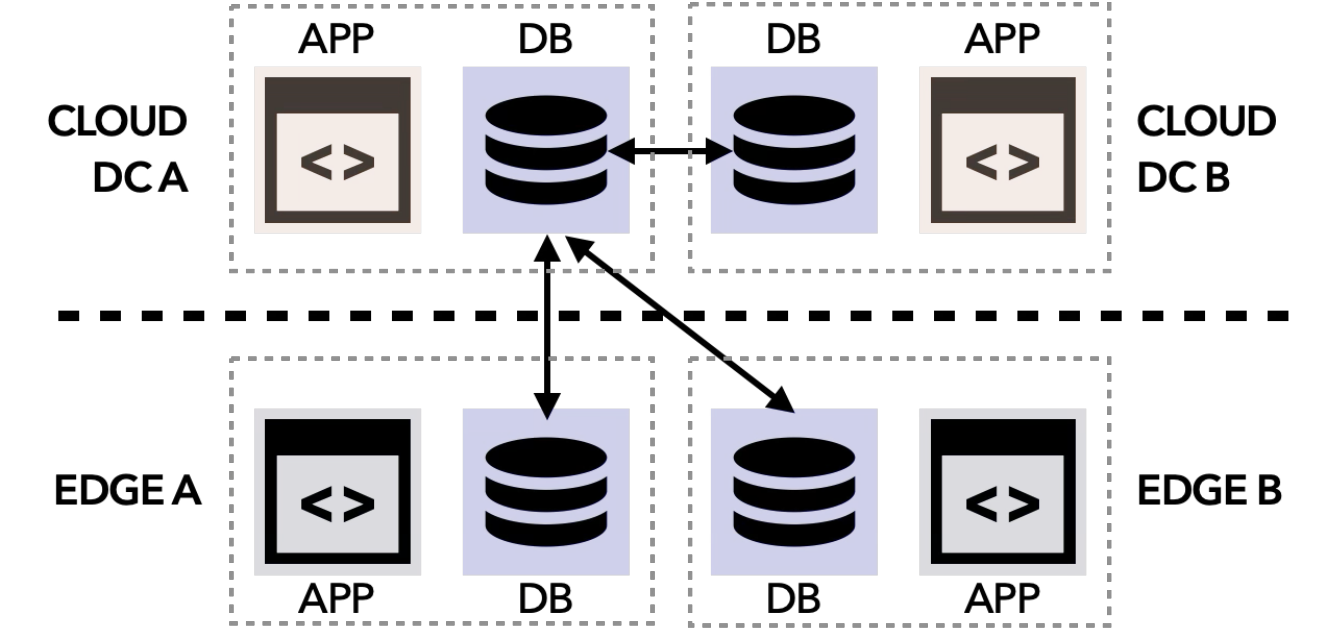}
         \subcaption{%
            Distributed Deployment
         }
         \label{fig: deployment distributed}
    \end{subfigure}
    \hfill
    \begin{subfigure}[t]{.48\linewidth}
         \centering
         \includegraphics[width=\linewidth]{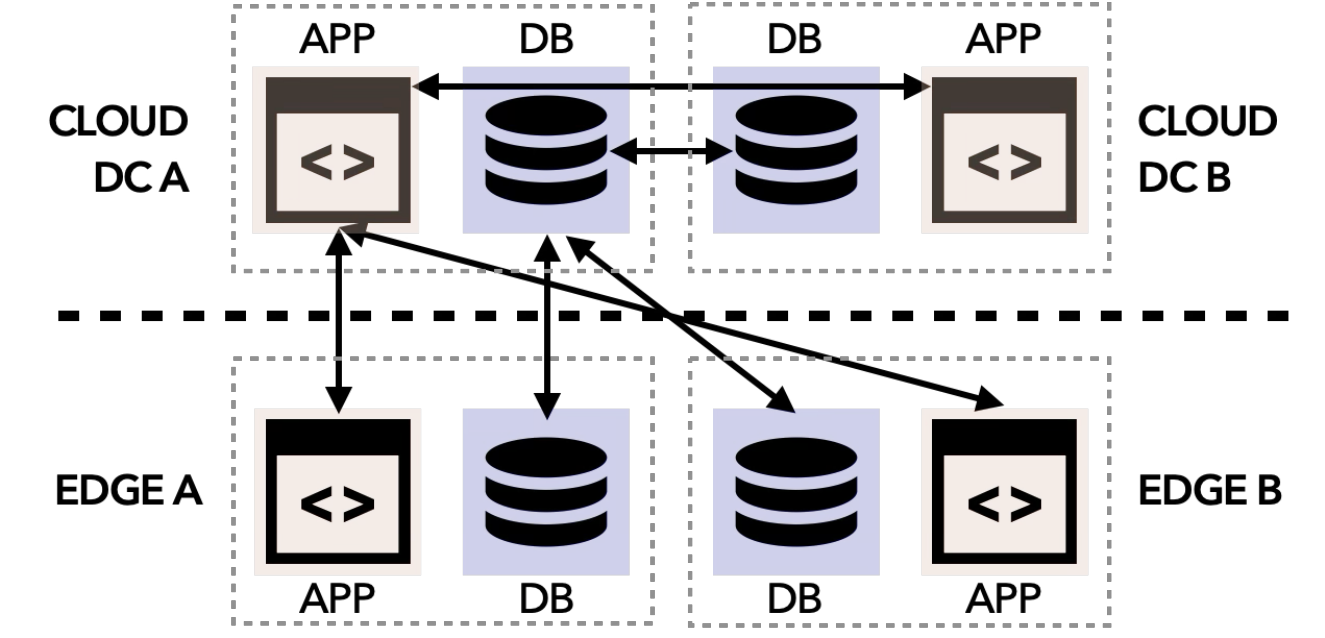}
         \subcaption{%
            Decentralized Deployment
         }
         \label{fig: deployment decentralized}
    \end{subfigure}
    \caption{%
        \textbf{Island Readiness of Application Deployment Strategies.}
        The background color indicates \hlbeige{stateless components}, \hlblue{stateful components}, and \hlgray{standby replicas.}
    }
    \label{fig: deployment strategies}
\end{figure*}

Section~\ref{sec: roadmap} describes the complete decentralization of systems to achieve island-ready application setups with at least one replica in every region.
Such deployments have a full replica of the application in each region, facilitating users on the island access to the application despite being disconnected from the outside world, i.e., there is no connection to the cloud or other regions.
However, full decentralization increases the complexity and operational costs of applications.

Figure~\ref{fig: deployment strategies} compares the traditional (centralized), highly available, distributed, and decentralized deployment of an application, showing which parts of the deployment would be available regionally during island scenarios.
The example demonstrates the deployment of applications with stateless and stateful components and considers an application server (stateless) and a database server (stateful).

\textit{Centralized deployments} (Figure~\ref{fig: deployment centralized}) without regional availability of the application are neither resilient against outages nor island-ready.
\hlbeige{Users connect to the application server} in normal operation.
The \hlblue{database server is co-located with the application} in the same data center.
In island operation, the application is not available.
This is how typical small applications are deployed, which is why they would not be available on islands.

\textit{High-availability deployments} (Figure~\ref{fig: deployment high-available}) with replicas of the application server and database in multiple clouds are scalable to billions of users.
However, such deployments offer no regional availability and would not be available on islands.
\hlbeige{Users connect to the closest cloud application server} in normal operation.
The \hlblue{database server is synchronized between clouds.}
In island operation, the application is not available.

\textit{Distributed deployments} (Figure~\ref{fig: deployment distributed}) with a distributed database and a standby application replica are available on islands, but only after manual transition.
Such deployments would be available on islands after the standby application replica is brought up.
\hlbeige{Users connect to the cloud application server} in normal operation.
The \hlblue{database server replicas synchronize to the main cloud server.}
In island operation, the \hlgray{regional standby application replica is activated.}

\textit{Decentralized deployments} (Figure~\ref{fig: deployment decentralized}) with synchronization of application and database replicas are island-ready.
Applications deployed like that could seamlessly transition to island operation without downtimes.
\hlbeige{Users connect to the regional application server} in normal and in island operation.
\hlblue{The application and database replicas synchronize to the main cloud servers.}
In island operation, synchronization is on hold, but the application is available.

Centralized and high-availability deployments are not island-ready as they lack regional availability of the application.
Distributed and decentralized deployments both support island scenarios, but only the decentralized approach can seamlessly transition into island operation.

The advantage of decentralization is that the system is island-ready by desig,n and island connectivity is available instantaneously, which might be crucial for crisis response.
The distributed approach requires activation of the standby replicas, which requires qualified staff in the region or automatic mechanisms to detect and carry out the transition into island operation.
On the other hand, a benefit of the distributed approach is that it preserves resources.
When an application's demand is insufficient to justify dedicated replicas in every region during normal operation, the stateless components can remain in standby, and only the stateful components are synchronized.

\subsection{Future Work}
We define island readiness and island-ready 6G communication networks (Section~\ref{sec: vision}) and identify involved actors and their use cases (Section~\ref{sec: perspectives}).
We evaluate the island readiness of state-of-the-art communication networks based on 5G and 5G-Advanced (Section~\ref{sec: architecture}) and outline which challenges remain open to achieve island-readiness in 6G (Section~\ref{sec: roadmap}).
Solving the described open challenges is beyond the scope of this work.
We believe that a participative and all-society approach bringing together all the mentioned stakeholders is required to address the challenges outlined in this article.


\section{Conclusion}
\label{sec: conclusion}

This article analyzed the island readiness of state-of-the-art cellular networks.
The resilience of today's cellular networks is insufficient to operate in crisis situations.
We identified users, mobile network operators, cloud and edge providers, and application developers as the leading actors of island connectivity.
We derived use cases for each actor and evaluated to what extent 5G and 5G-Advanced systems can support them.
Many use cases cannot be supported when the heavily involved core network is unavailable on an island.
Additionally, developers and providers must adapt their application architectures to become island-ready or develop recovery mechanisms that are applied when transitioning into island operation.
The main objective of this article was to identify relevant perspectives on island-ready 6G communication.
We encourage future work to deep dive into the challenges of individual actors, complementing the user perspective~\cite{janzen_2025_user} with perspectives of operators, developers, providers, and authorities.

\section*{Acknowledgements}
This work was supported by the Federal Ministry of Education and Research of Germany in the project Open6GHub (grant number: 16KISK014) and by the LOEWE initiative (Hesse, Germany) within the emergenCITY center [LOEWE/1/12/519/03/05.001(0016)/72].
We thank Lea Holaus for Figure~\ref{fig: vision}.

\bibliographystyle{elsarticle-num} 
\bibliography{references}

\end{document}